\documentclass[12pt,a4]{article}

\usepackage{color,tikz}
\usepackage[unicode,bookmarks,bookmarksopen,bookmarksopenlevel=2,colorlinks,linkcolor=blue,citecolor=red]{hyperref}

\usepackage[font=small,labelfont=bf]{caption}
\usepackage{booktabs}

\usepackage{amsmath,dsfont,eucal,amssymb,amsthm,amsfonts}
\usepackage{mathrsfs,graphicx,texdraw}

\newtheorem{example}{Example}
\newtheorem{remark}{Remark}

\DeclareMathOperator{\diag}{diag}

\def\openone{\leavevmode\hbox{\small1\kern-3.3pt\normalsize1}}

\def\diag{\mbox{diag}\,}

\newcommand{\rJ}{\mathrm{J}}
\newcommand{\re}{\mathrm{e}}
\newcommand{\ri}{\mathrm{i}}
\newcommand{\rd}{\mathrm{d}}

\usepackage{makeidx}
\usepackage{times}
\usepackage{calc}
\usepackage{xcolor}
\usepackage{titlesec}
\usepackage{bm}
\usepackage{stmaryrd}
\usepackage{bm}
\usepackage{framed}
\usepackage[T1]{fontenc}
\usepackage{pifont}
\usepackage{amsmath}
\usepackage{subfig}
\usepackage{dsfont}

\graphicspath{{figures/}} % Directory in which figures are stored
\begin{document}

\begin{center}
{\LARGE \bf On nonlocal reductions of the multi-component\\[6pt] nonlinear
Schr\"odinger  equation on symmetric spaces}

\bigskip

{\bf Georgi  G. Grahovski \footnote{E-mail: {\tt grah@essex.ac.uk}}, Junaid I. Mustafa \footnote{E-mail: {\tt jimust@essex.ac.uk}}
 and
Hadi Susanto \footnote{E-mail: {\tt  hsusanto@essex.ac.uk}}}

\end{center}

%\medskip

\begin{center}
{\it  Department of Mathematical Sciences, University of Essex,
 Wivenhoe Park, Colchester, UK }\\
\end{center}

\begin{abstract}
\noindent
The aim of this paper is to develop the inverse scattering transform (IST) for  multi-component generalisations of nonlocal reductions of the  nonlinear Schr\"odinger (NLS) equation with ${\cal PT}$-symmetry related to symmetric spaces. This includes: the spectral properties of the associated Lax operator, Jost function, the scattering matrix and the minimal set of scattering data, the fundamental analytic solutions.  As main examples we use the Manakov vector Schr\"odinger equation (related to {\bf A.III}-symmetric spaces) and the multi-component NLS (MNLS) equations of Kullish-Sklyanin type (related to {\bf BD.I}-symmetric spaces). Furthermore, the $1$- and $2$-soliton solutions are obtained by using an appropriate modification of the  Zakharov-Shabat dressing method. It is shown, that the MNLS equations of these types allow both regular and singular soliton configurations. Finally, we present here different examples of 1- and 2-soliton solutions for both types of models, subject to different reductions.
\end{abstract}

%$\bf{Keywords:}$ integrable system, MNLS equations, Lax representation, ZS system, spectral decompositions, ${\cal PT}$-symmetry, IST, Riemann-Hilbert problem, dressing method, %soliton solution, symmetric spaces, local and nonlocal involutions.

%\newpage
%\tableofcontents
%\newpage

\section{Introduction}
One of the most important and popular completely integrable nonlinear PDE is the nonlinear Schr\"{o}\-din\-ger (NLS) equation \cite{ZaSh*74a,FaTa,ZMNP,GVYa*08}:
\begin{equation}\label{Introduction1}
{\rm i}q_t+ q_{xx}+2|q|^2 q(x,t)=0.
\end{equation}
Here $q(x,t)$ is a complex valued function, tending fast enough to zero as $\left|x\right|\rightarrow \infty$ \cite{GVYa*08, AblPriTru}.
It  has  been  derived as a governing equation describing processes and phenomena  in  such  diverse  fields  as  deep  water  waves,  plasma  physics
and nonlinear fibre optics.  For instance, in optics, NLS models wave propagation in Kerr media, where
the nonlinearity is proportional to the intensity of the field.

The NLS equation \eqref{Introduction1} appeared at the early stage of the development of the inverse scattering method (ISM) and the theory of solitons \cite{AblPriTru, DraJoh, FaTa,GVYa*08, ZMNP}, and exhibits all remarkable properties of PDEs and systems of PDEs, integrable by the ISM: it allows soliton solutions, infinite set of integrals of motion, multi-Hamiltonian formulation and so on \cite{GVYa*08,FaTa}. The key tool in studying integrability by the ISM is the existence of Lax representation of the nonlinear evolutionary equation (NLEE), that is, the NLEE can be represented as a compatibility condition of two linear operators \cite{ZMNP,FaTa,ZaSh*74a,GVYa*08,IP2}. The scattering problem for \eqref{Introduction1} is given by the Zakharov-Shabat system (related to $sl(2,{\Bbb C})$ algebra):
\begin{align}\label{(ZS) System}
L\chi\equiv & \left({\rm i}\frac{d}{dx}+q(x,t)-\lambda\sigma_3\right)\chi\left(x,t,\lambda\right)=0,\nonumber
\\
q(x,t)= & \left(
\begin{array}{cc}
0 & q^+ \\
 q^- & 0\\
\end{array}
\right), \qquad
\sigma_3=\left(
\begin{array}{cc}
1 & 0\\
0 & -1\\
\end{array}
\right),
\end{align}
The first integrable multi-component generalisation of the scalar NLS equation (\ref{Introduction1}) is the Manakov vector NLS equation:
\begin{eqnarray}\label{Examples of MNLS-type equations4}
{\rm i}q_{1,t}+q_{1,xx}+2 (|q_1|^2+|q_2|^2)q_1(x,t)=0;\nonumber\\
{\rm i}q_{2,t}+q_{2,xx}+2 (|q_1|^2+|q_2|^2)q_2(x,t)=0.
\end{eqnarray}
Here, $ q(x,t)$ is a $2$-component vector-valued function. It is associated with a scattering problem of  Zakharov-Shabat type related to the algebra $sl(3,{\Bbb C})$.
It was proposed by S. V. Manakov \cite{Manakov}  as an asymptotic model for the propagation of the electric
field in a waveguide.   Subsequently, this system was derived as a key model for light
propagation in optical fibres \cite{Menyuk,AckMil}.

The Manakov model \eqref{Examples of MNLS-type equations4} can be generalised to $n$-component vectors (see \cite{ZMNP})
\begin{equation}\label{eq:Man}
i{\bf v}_t+{\bf v}_{xx}+2 ({\bf v}^\dag , {\bf v}) {\bf v}=0,\qquad {\bf v}= {\bf v}(x,t).
\end{equation}
Here ${\bf v}$ is an $n $-component complex-valued vector and $(\cdot ,\cdot) $ is the standard scalar product. It is again integrable by the ISM \cite{AblPriTru,ZMNP,FaTa,GVYa*08}.

The classical ZS system can be generalised to a matrix (multi-component) form in a number of  ways. One of the standard ways for doing this is by considering Lax operators taking values in a simple Lie algebra $\mathfrak{g}$:
\begin{align}\label{Matrix generalisations of NLS Hierarchies1}
L=\ri \partial_x + Q -\lambda \rJ.
\end{align}
Here, $\rJ$ is a constant element of simple Lie algebra $\mathfrak{g}$ \cite{Helg, Gilmore}. Generalizations of the NLS equation to symmetric spaces related to Lie algebra $\mathfrak{g}$ are proposed in \cite{ForKu*83,AthFor,For}.  This includes coupled NLS systems with Lax pairs related the symmetric spaces {\bf A.III}, {\bf C.I}, {\bf D.III} and {\bf BD.I} types in the Cartan classification \cite{Helg,loos}. Among this class, one can get as a special case the Manakov vector NLS equation and the one studied by Kulish and Sklyanin \cite{KulSkl}. All these generalizations are solvable by the ISM \cite{KulSkl,ForKu*83,GVYa*08}.

When the rank $r$ of the underlying simple Lie algebra ${\frak g}$ grows, the corresponding generic NLEE (or  systems) will contain as many independent complex-valued functions as the number of all roots of ${\frak g}$ \cite{DriSok,58,vgrn}. They are solvable for any $r $ but their possible applications
to physics for large $r $ do not seem realistic. However one still may
extract new integrable and physically useful NLEE by imposing reductions
on $L(t) $,  i.e.  algebraic restrictions on $Q(x,t) $ which diminish the
number of independent functions in them  and the number of equations
\cite{2}.  Of course, such restrictions must  be compatible with the
dynamics of the NLEE \cite{58,vgrn}.

Recently, in \cite{AblMus} was proposed  a nonlocal integrable equation of nonlinear Schr\"odinger type
\begin{equation}\label{AblMus}
{\rm i}q_t+ q_{xx}+V(x,t) q(x,t)=0, \qquad V(x,t)=2q(x,t)q^*(-x,t).
\end{equation}
with
${\cal PT}$-symmetry, due to the invariance of
the so-called self-induced potential $V(x,t)$ under the combined action of parity and time
reversal symmetry. In the same paper, the 1-soliton solution for this model is derived and it was shown
that it develops  singularities in finite time. Soon after this, nonlocal ${\cal PT}$-symmetric generalisations are found for
 the Ablowitz-Ladik model in \cite{AblMus1}. All these models are integrable by the Inverse Scattering Method (ISM) \cite{AblMus2,GeSa,zm1,ZaSh*74a}.
Such nonlocal reductions of the NLS equation and its multi-component generalisations are of particular interest in regards to applications in ${\cal PT}$-symmetric optics, especially in developing of theory of electromagnetic waves in  artificial heterogenic media \cite{Konotop, Barash, Barash1, Nature}. For an up-to-date review, see for example \cite{UFN}.

Historically the first pseudo-hermitian hamiltonian with real spectrum is the ${\cal PT}$-symmetric one in \cite{Bender1}. Pseudo-hermiticity here means that
the Hamiltonian ${\cal H}$ commutes with the operators of spatial reflection ${\cal P}$ and time reversal ${\cal T}$: ${\cal P}{\cal T}{\cal H}={\cal H}{\cal P}{\cal T}$ \cite{Ali1, Ali2}.
The action of these operators is defined as follows: ${\cal P}: x\to -x$ and ${\cal T}: t\to -t$ \cite{Bender2, Fring2}. Supposing that the wave function is a scalar, this leads to the following action of the operator of spatial reflection on the space of states: ${\cal P}\psi(x,t)=\psi(-x,t)$ and ${\cal T}\psi(x,t)=\psi^*(x,-t)$. As a result,  the Hamiltonian and the wave function are ${\cal PT}$-symmetric, if ${\cal H}(x,t)={\cal H}^*(-x,-t)$ and $\psi(x,t)=\psi^*(-x,-t)$ \cite{Fring1}. Here we used also, that the parity operator ${\cal P}$ is linear and unitary while the time reversal operator ${\cal T}$ is anti-linear and anti-unitary.

The aim of this paper is study nonlocal reductions and to derive the corresponding soliton solutions for multi-component NLS models related to symmetric spaces of {\bf A.III}-type and {\bf BD.I}-type. This will be done based on examples of the vector NLS equations related to symmetric spaces of ${\bf A.III}\simeq SU(s+p)/S(U(s)\otimes U(p))$ type and the Kulish-Sklyanin model related to symmetric spaces of  {\bf BD.I}$\simeq SO(2r+1)/SO(2)\otimes SO(2r-1)$ type.

The structure of this paper is as follows: In Section  2 we outline the form of the Lax operators and the general form of the NLEEs as well as the nonlocal symmetries (involutions) of interest. In Section 3 we present the direct scattering problem: the Jost solutions, the scattering matrix and the minimal set of scattering data and  the fundamental analytic solutions (FAS). Furthermore, In Section 4, based on appropriate modification of the Zakharov-Shabat dressing method, we derive 1- and 2-soliton solution of the corresponding NLEE equation related to ${\bf A.III}$ and ${\bf BD.I}$ symmetric spaces and study the effect on nonlocal reductions on them.

\section{Preliminaries}\label{sec:prelim}

\subsection{Lax pair and general form of the equations}\label{ssec:Lax}

Let us start with the generic Lax pair for the MNLS equations on symmetric spaces \cite{G2005}, it can be represented by the following form:
\begin{subequations}\label{Lax Representation g=so(n)}
\begin{align}
L\chi(x,t,\lambda)& \equiv \rm i \partial_{x} \chi+(Q(x,t)-\lambda \rJ)\chi (x,t,\lambda) =0,\quad U(x,t,\lambda)=Q(x,t)-\lambda \rJ,\label{Lax Representation g=so(n)a}
\\
M\chi(x,t,\lambda)& \equiv \rm i \partial_{t} \chi +(V_{0}(x,t)+\lambda V_{1}(x,t)- \lambda^{2} \rJ)\chi(x,t,\lambda)=0,
\label{Lax Representation g=so(n)b}
\\
V_{1}(x,t)& =  Q(x,t), \quad  V_{0}(x,t)= \ri \text{ad}^{-1}_{\rJ}\frac{\rd Q}{\rd x} +\frac{1}{2}[\text{ad}^{-1}_{\rJ} Q,Q(x,t)].\label{Lax Representation g=so(n)c}
\end{align}
\end{subequations}
Here $\chi (x,t,\lambda)$ are the eigenfunction of the Lax operators, $U(x,t,\lambda)$ and $V(x,t,\lambda)$ are chosen to take values in a simple  Lie algebra $\mathfrak{g}$ of rank $r$, while the eigenfunctions $\chi(x,t,\lambda)$ belong to the corresponding Lie group $\mathfrak{G}$. Here also $\rJ$ is  a constant element of the Cartan subalgebra $\mathfrak{h}\subset \mathfrak{g}$ (which can be always chosen to be represented by a diagonal matrix) and $\lambda\in \mathbb{C}$ is a spectral parameter. On the  ${\bf A.III} \simeq SU(p+s)/S(U(p)\times U(s))$ symmetric spaces the potential of $L(\lambda)$ \eqref{Lax Representation g=so(n)a} reads:
\begin{align}
Q(x,t)= &
\left(\begin{array}{cc}
0 & \mathbf{q}^+ \\
\mathbf{q}^- & 0 \\
\end{array}
\right),
\hspace{0.3in}
\rJ=\frac{2}{s+p}
\left(
  \begin{array}{cc}
    p \mathds{1}_s & 0 \\
    0 & -s\mathds{1}_p \\
  \end{array}
\right),\label{Lax representation1d}
\end{align}
where $\hm{q}^+(x,t)$ and $(\hm{q}^-)^T(x,t)$ are $s \times p$ matrix-valued functions belonging to the simple Lie algebra $\mathfrak{g}$, $\mathds{1}_s$ and $\mathds{1}_p$ are the $s\times s$ and $p\times p$ identity matrices, respectively, $s+p=n$.

The NLEE can be written as a compatibility condition
\begin{equation}\label{Generalizations of the AKNS Approach1}
[L(\lambda),M(\lambda)]=0,
\end{equation}
of the two Lax operators \eqref{Lax Representation g=so(n)a}-\eqref{Lax Representation g=so(n)b}. In particular, if $L(\lambda)$ and $M(\lambda)$ are related to $ {\bf A.III}\simeq SU(p+1)/S(U(1)\otimes U(p))$ - symmetric spaces \cite{ForKu*83}, the explicit parametrisation \eqref{Lax representation1d} of the ${\bf A.III}$ symmetric spaces will give the system
\begin{eqnarray}\label{Examples of MNLS-type equations4a}
{\rm i}\mathbf{q}^+_{t}+\mathbf{q}^+_{xx}+2 \mathbf{q}^+\mathbf{q}^-\mathbf{q}^+(x,t)&=&0,\\
-{\rm i}\mathbf{q}^-_{t}+\mathbf{q}^-_{xx}+2 \mathbf{q}^-\mathbf{q}^+\mathbf{q}^-(x,t)&=&0.
\end{eqnarray}
The particular choice $s=1$ and $p=2$ (assuming also the standard involution ${\bf q}^- = ({\bf q}^+)^*$ for the NLS type of models) corresponds to the well-know Manakov  system \cite{Manakov}. Its generalisations for $n$-dimensional vectors ${\bf q}^{\pm}$ is known as vector nonlinear Schr\"odinger equation \cite{GVYa*08}.

Another class of the multi-component NLS (MNLS) equations is known as Kulish and Sklyanin (KS) models and are related to the symmetric spaces of {\bf BD.I}$\simeq SO(2r+1)/SO(2)\otimes SO(2r-1)$ type \cite{GKKV*08, G37-52, G2017, GG2010, G2009}.  The generic NLEEs of this class can be written as:
\begin{align}\label{Some Algebraic Properties for Cartan Weyl Basis5}
\ri\vec{\mathbf{q}}_{t}^{+}+\vec{\mathbf{q}}_{xx}^{+}+ 2\left(\vec{\mathbf{q}}^{+},\vec{\mathbf{q}}^{-}\right) \vec{\mathbf{q}}^{+}- \left(\vec{\mathbf{q}}^{+},s_{0}\vec{\mathbf{q}}^{+} \right)s_{0}\vec{\mathbf{q}}^{-}= &0,\nonumber
\\
\ri\vec{\mathbf{q}}_{t}^{-}-\vec{\mathbf{q}}_{xx}^{-}- 2\left(\vec{\mathbf{q}}^{+},\vec{\mathbf{q}}^{-}\right) \vec{\mathbf{q}}^{-}+ \left(\vec{\mathbf{q}}^{-},s_{0}\vec{\mathbf{q}}^{-} \right)s_{0}\vec{\mathbf{q}}^{+}=&0,
\end{align}
which is associated with ${\mathfrak{g}\simeq so(2r+1,{\Bbb C})}$  linear system \eqref{Lax Representation g=so(n)a} where
\begin{align}\label{Some Algebraic Properties for Cartan Weyl Basis6}
Q(x,t)=
\left(
  \begin{array}{ccc}
    0 & \mathbf{(q^{+})^{T}} & 0 \\
    \mathbf{q^{-}} &{\bf 0} & s_{0}\mathbf{q^{+}} \\
    0& \mathbf{(q^{-})^{T}}s_{0} & 0 \\
  \end{array}
\right), \qquad \rJ =\diag (1, {\bf 0},-1).
\end{align}
Here $\mathbf{q^{\pm}}$  are $2r-1$-component vectors, and  $s_{0}$ is the matrix that defines the orthogonal algebra ${\bf B}_r\simeq so(2r+1)$:
\begin{align}\label{Lax Representation g=so(n)3}
X\in so(2r+1) \ \ \text{iff} \ \  X+S_{0}X^{T}S_{0}^{-1}=0, \quad S_{0}=\sum_{s=1}^{2r+1}(-1)^{s+1}E_{s,2r+2-s}^{(2r+1)}= \left(
  \begin{array}{ccc}
    0 &0 & 1 \\
    0 &-s_{0} & 0 \\
    1& 0 & 0 \\
  \end{array}
\right),
\end{align}
where $E_{s,p}^{(2r+1)}$ is a $2r+1 \times 2r+1$ matrix with matrix elements given by $(E_{s,p}^{(2r+1)})_{ij}= \delta_{si}\delta_{pj}$.

\subsection{Symmetries of the Lax operator and reductions}\label{Symmetries of the Lax operator and reductions}

A systematic way of description and classification of the class of admissible reductions for a given Lax pair is introduced  by A. Mikhailov in \cite{2}. This is done by the so-called reduction group. The reduction group is a finite group which preserves the Lax representation, i.e. it ensures that the reduction constraints are automatically compatible with the evolution \cite{58, g2003, vgrn, TV}. The reduction constrains diminish the number of independent functions and the number of equations. Here we will restrict ourselves with $\mathbb{Z}_2$ reductions of the following two types:
\begin{align}\label{The Local Involutions2m}
\mbox{(A)} &&\qquad U(x,t,\lambda) = B (U(x,t,\lambda^{\ast}))^{\dag}B^{-1},
\\
\mbox{(B)} &&\qquad -U(x,t,\lambda) =B (U(-x,t,-\lambda^{\ast}))^{\dag}B^{-1},\label{The Local Involutions2m1}
\end{align}
where $B$ is constant block-diagonal matrix of the form $B = \left(
\begin{array}{cc}
\mathbf{B_+}  &0 \\
0& \mathbf{B_-} \\
\end{array}\right)$ for ${\bf A.III}$ symmetric spaces. For symmetric spaces of type ${\bf BD.I}$ it can be written in the form:
\begin{align}\label{The Local Involutions3mKS}
B = \left(
\begin{array}{ccc}
\mathbf{B_+ } &0 &0 \\
0& \mathbf{B_\pm}&0 \\
0 &0 &\mathbf{B_- } \\
\end{array}\right).
\end{align}
Here we assume also that the blocks $\mathbf{B_+}$ and $\mathbf{B_-}$ are nonsingular matrices.  Note also, that the involution of type (B) in (\ref{The Local Involutions2m1}) results in nonlocal reduction conditions.

Recalling the standard block-structure on ${\bf A.III}$ symmetric spaces (\ref{Lax representation1d}) one can write explicitly the reductions conditions for the matrix blocks:
\begin{align}\label{The Local Involutions4m}
\mbox{(A)} &&\qquad\mathbf{q}^{-}(x,t)=\mathbf{B_-}(\mathbf{q}^{+}(x,t))^{\dag}(\mathbf{B_+})^{-1},\quad \mathbf{q}^{+}(x,t)=\mathbf{B_+}(\mathbf{q}^{-}(x,t))^{\dag}(\mathbf{B_-})^{-1},\\
\mbox{(B)} &&\qquad\label{The Nonlocal Involution4m}
\mathbf{q}^{-}(x,t)=-\mathbf{B_-}(\mathbf{q}^{+}(-x,t))^{\dag}(\mathbf{B_+})^{-1},\quad \mathbf{q}^{+}(x,t)=-\mathbf{B_+}(\mathbf{q}^{-}(-x,t))^{\dag}(\mathbf{B_-})^{-1}.
\end{align}
Using the the local and nonlocal involutions \eqref{The Local Involutions4m} and \eqref{The Nonlocal Involution4m} respectively, one can write the resulting  Kulish-Sklyanin models. The local involution\eqref{The Local Involutions4m} will give  the following equations:
\begin{equation}\label{The Local and Nonlocal (KS) type Models VNLS Equation1}
\ri \mathbf{q}_{t}^{+}+\mathbf{q}_{xx}^{+}+2\left(\mathbf{q}^{+}(x,t),(\mathbf{q}^{+}(x,t))^{\ast}\right)\mathbf{q}^{+}-\left(\mathbf{q}^{+}, s_{0}\mathbf{q}^{+}\right)s_{0}(\mathbf{q}^{+}(x,t))^{\ast}=0,
\end{equation}
while the nonlocal involution \eqref{The Nonlocal Involution4m} will result in:
\begin{equation}\label{The Local and Nonlocal (KS) type Models VNLS Equation2}
\ri \mathbf{q}_{t}^{+}+\mathbf{q}_{xx}^{+}+2\left(\mathbf{q}^{+}(x,t),(\mathbf{q}^{+}(-x,t))^{\ast}\right)\mathbf{q}^{+}-\left(\mathbf{q}^{+}, s_{0}\mathbf{q}^{+}\right)s_{0}(\mathbf{q}^{+}(-x,t))^{\ast}=0.
\end{equation}

\noindent
{\bf Example A. [Manakov model]} {\it If ${\frak g}\simeq sl(3,{\Bbb C})$,  $p=1$ and $s=2$, then $\hm{q}^+(x,t)$ and $\hm{q}^-(x,t)$ are 2-component vector functions. In addition, If we apply a (local) reduction of type (A) with $B={\Bbb I}$, this will reproduce the standard Manakov VNLS equation:
\begin{equation}\label{The Nonlocal MNLS Equation1}
\begin{array}{cc}
-{\rm i}q_{1,t}+ q_{1,xx}+2 \ (|q_{1}(x,t)|^2+ |q_{2}(x,t)|^2) \ q_{1}(x,t)=0,
\vspace{0.1in}\\
-{\rm i}q_{2,t}+ q_{2,xx}+2 \ (|q_{1}(x,t)|^2+ |q_{2}(x,t)|^2) \  q_{2}(x,t)=0.
\end{array}
\end{equation}
Taking an involution of type (B) with $\mathbf{B}_+=\mp1$ and $\mathbf{B}_-=\text{diag}(\pm1, \pm1)$, then we obtain the following nonlocal reduction of  Manakov model:}
\begin{equation}\label{The Nonlocal MNLS Equation3}
\begin{array}{cc}
-{\rm i}q_{1,t}+ q_{1,xx}+2 \ (q_{1}(x,t) \ q_{1}^{\ast}(-x,t)+ q_{2}(x,t) \ q_{2}^{\ast}(-x,t)) \ q_{1}(x,t)=0,
\vspace{0.1in}\\
-{\rm i}q_{2,t}+ q_{2,xx}+2 \ (q_{1}(x,t) \ q_{1}^{\ast}(-x,t)+ q_{2}(x,t) \ q_{2}^{\ast}(-x,t)) \  q_{2}(x,t)=0.
\end{array}
\end{equation}

\noindent
{\bf Example B. [Kulish-Sklyanin model]} {\it In the simplest case of Lax operators related to $SO(5)/SO(2)\otimes SO(3)$ (${\rm rank}\, {\frak g}=2$), we can set ${\bf q}^+=(q_{12}^+,q_{13}^+,q_{14}^+)$ and ${\bf q}^-=(q_{12}^-,q_{13}^-,q_{14}^-)$. After assuming also the standard involution of type (A) with ${B}={\Bbb I}$,  the compatibility condition (\ref{Generalizations of the AKNS Approach1}) will lead to the following  3-component NLS system:
\begin{align}\label{Some Algebraic Properties for Cartan Weyl Basis7}
\ri(q_{t}^{+})_{12}&+(q_{xx}^{+})_{12}+ 2\left(\left|q_{12}\right|^{2}+2\left|q_{13}\right|^{2}\right) q_{12} +2(q^{+}_{14})^{*}(q^{+}_{13})^{2}=0 ,\nonumber
\\
\ri(q_{t}^{+})_{13}&+(q_{xx}^{+})_{13}+ 2\left(\left|q_{12}\right|^{2}+\left|q_{13}\right|^{2} +\left|q_{14}\right|^{2}\right) q_{13} +2(q^{+}_{13})^{*}q^{+}_{14}q^{+}_{12}=0 ,\nonumber
\\
\ri(q_{t}^{+})_{14}&+(q_{xx}^{+})_{14}+ 2\left(\left|q_{14}\right|^{2}+2\left|q_{13}\right|^{2}\right) q_{14} +2(q^{+}_{12})^{*}(q^{+}_{13})^{2}=0.
\end{align}
This appears to be a model describing ${\cal F}=1$ spinor Bose-Einstein condensates in one-dimensional approximation \cite{KulSkl,G2009,AGGK,GGK2005c}.
}

\section{Direct Scattering Transform for $L(\lambda)$}\label{sec:DST}

\subsection{Jost solutions and scattering matrix}

The starting point here are the so-called Jost solutions, which are determined by their asymptotics for $\left|x\right|\rightarrow \infty$:
\begin{equation}\label{Jost Functions2}
\lim_{x\rightarrow \infty}\re^{{\rm i}\lambda \rJ x}\psi(x,t,\lambda)=\mathds{1}_{s+p},\qquad \lim_{x\rightarrow -\infty}\re^{{\rm i}\lambda \rJ x}\phi(x,t,\lambda)=\mathds{1}_{s+p}, \qquad \lambda \in \mathbb{R}.
\end{equation}
Along with these functions, one can also use "normalised to unity" Jost solutions
\begin{equation}\label{Jost Functions6}
\xi(x,t,\lambda)=\psi(x,t,\lambda)\re^{({\rm i}\lambda \rJ x)},\qquad
\varphi(x,t,\lambda)=\phi(x,t,\lambda)\re^{({\rm i}\lambda \rJ x)},
\end{equation}
satisfying the following equation
\begin{equation}\label{Jost Functions8}
{\rm i}\frac{\rd \xi}{\rd x}+Q(x,t)\xi(x,t,\lambda)-\lambda\left[\rJ,\xi(x,t,\lambda)\right]=0,
\end{equation}
provided that \eqref{Jost Functions2} satisfy \eqref{Lax Representation g=so(n)a}.

On the continuous spectrum of $L(\lambda)$ the two Jost solutions $\psi(x,t,\lambda)$ and $\phi(x,t,\lambda)$ are related via scattering matrix $T(\lambda)$
\begin{equation}\label{Scattering Data4}
\phi(x,t,\lambda)=\psi(x,t,\lambda)T(\lambda), \qquad \lambda\in\mathbb{R}.
\end{equation}
The scattering matrix $T(\lambda)$ belongs to the Lie group ${\frak G}$ corresponding the Lie algebra /symmetric space of $L(\lambda)$. For symmetric spaces of type ${\bf A.III}$ in the Cartan classification, it has the following block structure:
\begin{equation}\label{Scattering Data6}
T(\lambda)=\left(\begin{array}{cc}
      \hm{a}^+(\lambda) & -\hm{b}^-(\lambda) \\
      \hm{b}^+(\lambda) & \hm{a}^-(\lambda)
    \end{array}\right),
\end{equation}
where, $\hm{a}^+(\lambda)$ and $\hm{a}^-(\lambda)$ are square matrices, while $\hm{b}^+(\lambda)$ and $\hm{b}^-(\lambda)$ are rectangular matrices. The block structure of the inverse of $T(\lambda)$ can be written as:
\begin{align}\label{Scattering Data25}
\hat{T}(\lambda)=\left(\begin{array}{cc}
      \hm{c}^-(\lambda) & \hm{d}^-(\lambda) \\
      -\hm{d}^+(\lambda) & \hm{c}^+(\lambda)
    \end{array}\right),
\qquad \hat{T}(\lambda)= T^{-1}(\lambda),
\end{align}
where
\begin{align}\label{Scattering Data27}
\begin{array}{cc}
\hm{c}^\pm(\lambda)=\hat{\hm{a}}^\mp(\lambda)(\mathds{1}+\rho^{\pm}\rho^{\mp})^{-1}=  (\mathds{1}+\tau^{\mp}\tau^{\pm})^{-1}\hat{\hm{a}}^{\mp}(\lambda),
\\
\hspace{0.8in} \hm{d}^\pm(\lambda)=\hat{\hm{a}}^\mp(\lambda) \rho^{\pm}(\lambda)(\mathds{1}+\rho^{\mp}\rho^{\pm})^{-1}= (\mathds{1}+\tau^{\mp}\tau^{\pm})^{-1}\tau^{\mp}(\lambda)\hat{\hm{a}}^{\pm}(\lambda).
\end{array}
\end{align}
Here  $\rho^{\pm}(\lambda)$ and $\tau^{\pm}(\lambda)$ are the reflection and transmission coefficients respectively:
\begin{align}\label{Scattering Data28}
\rho^{\pm}(\lambda)=\mathbf{b}^{\pm}\hat{\mathbf{a}}^{\pm}(\lambda) =\hat{\mathbf{c}}^{\pm}\mathbf{d}^{\pm}(\lambda),\qquad \tau^{\pm}(\lambda)=\hat{\mathbf{a}}^{\pm}\mathbf{b}^{\mp}(\lambda) =\mathbf{d}^{\mp}\hat{\mathbf{c}}^{\pm}(\lambda).
\end{align}
For symmetric spaces of ${\bf BD.I}$ type, the block structure of $T(t,\lambda)$ and its inverse take the form:
\begin{align}\label{The Direct Scattering Transform for $L$1}
T(t,\lambda)=\left(
               \begin{array}{ccc}
                 m_{1}^{+} & -\vec{\mathbf{B}}^{- T} &c_{(1)}^{-} \\
                \vec{\mathbf{ b}}^{+} & \mathbf{T}_{22} &-s_{0}\vec{\mathbf{b}}^{-} \\
                 c_{(1)}^{+} & \vec{\mathbf{B}}^{+ T}s_0 &  m_{1}^{-} \\
               \end{array}
             \right), \qquad
\hat{T}(t,\lambda)=\left(
               \begin{array}{ccc}
                 m_{1}^{-} & \vec{\mathbf{b}}^{- T} &c_{(1)}^{-} \\
                 -\vec{\bf{B}}^{+} & \hat{\mathbf{T}}_{22} &s_{0}\vec{\bf{B}}^{-} \\
                 c_{(1)}^{+} & -\vec{\mathbf{b}}^{+ T}s_0 &  m_{1}^{+} \\
               \end{array}
             \right).
\end{align}
Here, $\vec{\bf{B}}^{\pm}(t,\lambda)$ and $\vec{\bf{b}}^{\pm}(t,\lambda)$ are $2r-1$-component vectors and $c_{(1)}^{\pm}(\lambda)$ and $m_{1}^{\pm}(\lambda)$ are scalar functions. The reflection coefficients $\vec{\rho}^\pm (\lambda)$, the transmission coefficients $\vec{\tau}^\pm (\lambda)$ and the functions $c^\pm_{(1)} (\lambda)$ are determined by the  generalised  Gauss decomposition  of $T(t,\lambda)$:
\begin{align}\label{The Direct Scattering Transform for $L$1a}
T(t,\lambda)= T^{-}_{\rJ}(t,\lambda)D^{+}_{\rJ}(t,\lambda)\hat{S}^{+}_{\rJ}(t,\lambda)= T^{+}_{\rJ}(t,\lambda)D^{-}_{\rJ}(t,\lambda)\hat{S}^{-}_{\rJ}(t,\lambda).
\end{align}
Here, $S^{\pm}_{\rJ}$ and $T^{\pm}_{\rJ}$ are upper- and lower- block-triangular matrices, which can be written in the form:
\begin{align}\label{The Direct Scattering Transform for $L$3}
S^{\pm}_{\rJ}=\re^{(\pm(\vec{\tau}^{\pm}(\lambda,t).\vec{E}^{\pm}_{1}))}, \qquad T^{\pm}_{\rJ}=\re^{(\mp(\vec{\rho}^{\mp}(\lambda,t).\vec{E}^{\pm}_{1}))},
\end{align}
where
\begin{align}\label{The Direct Scattering Transform for $L$4}
\vec{\tau}^{\pm}(\lambda,t)=\vec{\mathbf{B}}^{\mp}/m^{\pm}_{1}, \quad \vec{\rho}^{\pm}(\lambda,t)=\vec{\mathbf{b}}^{\pm}/m^{\pm}_{1}, \quad c_{(1)}^\pm=\frac{m_{1}^{\pm}}{2}(\vec{\rho}^{\pm,T}, s_{0}\vec{\rho}^{\pm})= \frac{m_{1}^{\mp}}{2}(\vec{\tau}^{\mp,T}, s_{0}\vec{\tau}^{\mp}),
\end{align}
and $D^{\pm}_{\rJ}(t,\lambda)$ is the block-diagonal factor in \eqref{The Direct Scattering Transform for $L$1a}:
\begin{align}\label{The Direct Scattering Transform for $L$5}
D^{+}_{\rJ}=\left(
              \begin{array}{ccc}
                m^{+}_{1} & 0 & 0 \\
                0 & \mathbf{m}_{2}^{+} & 0 \\
                0 & 0 & 1/m_{1}^{+}\\
              \end{array}
            \right),
\qquad
D^{-}_{\rJ}=\left(
              \begin{array}{ccc}
                1/m_{1}^{-} & 0 & 0 \\
                0 & \mathbf{m}_{2}^{-} & 0 \\
                0 & 0 & m_{1}^{-} \\
              \end{array}
            \right).
\end{align}
Here, $m_k^\pm(t,\lambda)$ are the upper/lower rank $k$ principal minors of the scattering matrix $T(t,\lambda)$ \eqref{The Direct Scattering Transform for $L$1a} and
\begin{align}\label{The Direct Scattering Transform for $L$6}
\mathbf{m}_{2}^{+}=\mathbf{T}_{22}+\frac{\vec{b}^{+}\vec{B}^{-,T}}{m^{+}_{1}}, \qquad   \mathbf{m}_{2}^{-}=\mathbf{T}_{22}+\frac{s_{0}\vec{B}^{-}\vec{b}^{+,T}s_{0}}{m^{-}_{1}}.
\end{align}
If the potential matrix $Q(x,t)$ satisfies the NLEE \eqref{Generalizations of the AKNS Approach1} then the associated scattering matrix $T(t,\lambda)$ evolves in time linearly, i.e. it satisfies the equation
\begin{equation}\label{eq:Scatt_t}
{\rm i}\ \frac{{\rm d}T}{{\rm d}t}+[{\rm f}(\lambda),T(t,\lambda)]=0.
\end{equation}
Here $f(\lambda)$ is the dispersion law of the NLEE \eqref{Generalizations of the AKNS Approach1}. For the NLS type of equations, we have $f(\lambda)=-\lambda^2 J$.

\subsection{Fundamental Analytic Solutions (FAS)}
In the this section, we will briefly outline the construction of the Fundamental Analytic Solutions (FAS) $\chi(x,,t, \lambda)$ of the generalised Zakharov-Shabat  system \eqref{Lax Representation g=so(n)} \cite{G2005, GVYa*08}. The FAS can be directly obtained from Jost solutions of \eqref{Lax Representation g=so(n)}:
\begin{align}\label{The Direct Scattering Transform for $L$2}
\chi^{\pm}(x,t,\lambda)=\phi (x,t,\lambda) S ^{\pm}_{\rJ}(t,\lambda)=\psi(x,t,\lambda)T^{\mp}_{\rJ}(t,\lambda) D^{\pm}_{\rJ}(t,\lambda),
\end{align}
by using the generalised Gauss decomposition \eqref{The Direct Scattering Transform for $L$1a} of the scattering matrix $T(t,\lambda)$.

On the real axis (i.e. on the continuous spectrum of $L(\lambda)$), the two FAS are linearly dependent:
 \begin{align}
\chi^{+}(x,t,\lambda)=&\chi^{-}(x,t,\lambda) G_0(t,\lambda),\qquad  \lambda\in \mathbb{R},\label{The Zakharov-Shabat Dressing Method Local1}
\end{align}
where the sewing function $G_0(t,\lambda)$ can be expressed in terms of the generalised Gauss factors $S ^{\pm}_{\rJ}(t,\lambda)$:
\begin{equation}\label{eq:Sewing}
G_{0,J}(t,\lambda)=\hat{S}_J^{-}(t,\lambda)S_J^{+}(t,\lambda)|_{t=0}.
\end{equation}
The independent matrix elements of $G_{0}(t,\lambda)$, together with the discrete spectrum of $L(\lambda)$ form up the  minimal set of scattering data of $L$. Based on the completeness relations of the associated square solution and Wronskian type of relation, one can recover the potential $Q(x,t)$ out of the minimal set of scattering data \cite{G*86}.

We end up this section with the remark that, although the general form of the Gauss decomposition \eqref{The Direct Scattering Transform for $L$1a} holds true for any symmetric space, for symmetric spaces of type ${\bf A.III}$ one can simplify the form of the matrix block by slightly modifying the Gauss decomposition of $T(\lambda)$ \footnote{Decomposition of type \eqref{the FAS16} are known as LU decomposition of $T(\lambda)$, while a decomposition of type \eqref{The Direct Scattering Transform for $L$1a} is known as LDU decomposition of $T(\lambda)$.}:
\begin{align}\label{the FAS16}
T(\lambda)= & \mathbf{T}_J^{-}(\lambda) \hat{\mathbf{S}}_J^{+}(\lambda) =\mathbf{T}^{+}_J(\lambda) \hat{\mathbf{S}}_J^{-}(\lambda).
\end{align}
Now, $\mathbf{S}^{\pm}(\lambda)$ and $\mathbf{T}^{\pm}(\lambda)$ are the block-triangular matrices:
\begin{equation}\label{the FAS15}
\begin{array}{cc}
\mathbf{S}_J^{+}(\lambda)=\left(\begin{array}{cc}
      \mathds{1}_s, & \hm{d}^-(\lambda) \\
      0, & \hm{c}^+(\lambda)
    \end{array}\right),\qquad
    \mathbf{T}_J^{-}(\lambda)=\left(\begin{array}{cc}
      \hm{a}^+(\lambda), & 0 \\
      \hm{b}^+(\lambda), & \mathds{1}_p
    \end{array}\right),
    \\
    \\
     \mathbf{S}_J^{-}(\lambda)=\left(\begin{array}{cc}
      \hm{c}^-(\lambda), & 0 \\
      -\hm{d}^+(\lambda), & \mathds{1}_p
    \end{array}\right),\qquad
    \mathbf{T}_J^{+}(\lambda)=\left(\begin{array}{cc}
      \mathds{1}_s, & -\hm{b}^-(\lambda) \\
      0, & \hm{a}^-(\lambda)
    \end{array}\right).
    \end{array}
\end{equation}
Using the Gauss decomposition \eqref{the FAS16} one can write explicitly the sewing function $G_{0,J}(t,\lambda)$ \eqref{eq:Sewing} in terms of the blocks of $T(t,\lambda)$ and it inverse:
\begin{align}
G_{0}(\lambda)= \hat{D}^{-}(\lambda)(\mathds{1}+K^{-}(\lambda)),\qquad \hat{G}_{0}(\lambda)= \hat{D}^{+}(\lambda)(\mathds{1}-K^{+}(\lambda)).\label{the FAS18}
\end{align}
Here  the block-diagonal factors $D^{\pm}(\lambda)$ and block-off-diagonal $K^{\pm}(\lambda)$ factors are expressed in terms of the blocks ${\bf a}^\pm (\lambda)$, ${\bf b}^\pm (\lambda)$ of $T(\lambda)$ \eqref{Scattering Data6} and in terms of the blocks ${\bf c}^\pm (\lambda)$, ${\bf d}^\pm (\lambda)$ \eqref{Scattering Data25} of  its inverse $\hat{T}(\lambda)$, respectively;
\begin{align}\label{the FAS19}
D^{-}(\lambda)= &\left(
  \begin{array}{cc}
    \mathbf{c}^{-}(\lambda) & 0 \\
    0 & \mathbf{a}^{-}(\lambda) \\
  \end{array}
\right),\qquad K^{-}(\lambda)=\left(
  \begin{array}{cc}
    0 & \mathbf{d}^{-}(\lambda)\\
    \mathbf{b}^{+}(\lambda) &0 \\
  \end{array}
\right),
\\
D^{+}(\lambda)= &\left(
  \begin{array}{cc}
    \mathbf{a}^{+}(\lambda) & 0 \\
    0 & \mathbf{c}^{+}(\lambda) \\
  \end{array}
\right),\qquad K^{+}(\lambda)=\left(
  \begin{array}{cc}
    0 & \mathbf{b}^{-}(\lambda)\\
    \mathbf{d}^{+}(\lambda) &0 \\
  \end{array}
\right).
\end{align}
The superscripts $``\pm''$ in the expressions $D^{\pm}(\lambda)$ above  mean analyticity for $\lambda \in {\Bbb C}_\pm$.

\section{Dressing method and soliton solutions}\label{sec:Dress}

\subsection{The Zakharov-Shabat Dressing Method}\label{ssec:ZSh-dress}

The main idea of  the Zakharov-Shabat dressing method is to start with a regular solution $\eta^{\pm}_{0}(x,t,\lambda)$ of  \eqref{Lax Representation g=so(n)a} and to perform a "dressing" by adding singularities at two prescribed points $\lambda_k^\pm$ \cite{RI, GKKV*08, 66, 67, G2005, ZMNP, Sh, DokLeb}. The new singular solution will have the form:
\begin{subequations}\label{The Dressing Method: The Singular SolutionsLocal2}
\begin{align}
\eta^{\pm}(x,t,\lambda)= & u_{k} (x,t,\lambda)\eta^{\pm}_{0}(x,t,\lambda)w_{k,\pm}^{-1}(\lambda).\label{The Dressing Method: The Singular SolutionsLocal2a}
\end{align}
\end{subequations}
Here, the matrices $w_{k,\pm}(\lambda)$ can be given by:
\begin{align}\label{The Dressing Method: The Singular SolutionsLocal20}
w_{k,+}(\lambda)=&\mathds{1},\qquad w_{k,-}(\lambda) = \left(
    \begin{array}{cc}
      u_{11,k}^{+} & 0 \\
      0 & u_{22,k}^{-} \\
    \end{array}
  \right),
\end{align}
where the limits of the dressing factor can be obtained by:
\begin{align}\label{The Dressing Method: The Singular SolutionsLocal19}
\lim_{x \rightarrow \infty}u_{k}(x,t,\lambda)=\left(
                                          \begin{array}{cc}
                                             u_{11,k}^{+} & 0 \\
                                            0 & \mathds{1}\\
                                          \end{array}
                                        \right), \qquad  u_{11,k}^{+} = \mathds{1}+ (c_{k}(\lambda)-1)P_{11, k}^{+},
                                        \\
\lim_{x \rightarrow -\infty}u_{k}(x,t,\lambda)=\left(
                                          \begin{array}{cc}
                                             \mathds{1}& 0 \\
                                            0 &  u_{22,k}^{-}\\
                                          \end{array}
                                        \right), \qquad  u_{22,k}^{-} = \mathds{1}+ (c_{k}(\lambda)-1)P_{22, k}^{-}.
\end{align}
For Lax operators relatred to ${\bf A.III}$ symmetric spaces, the dressing factors have the form:
\begin{equation}
u_{k} (x,t,\lambda)=  \mathds{1} + (c_{k}(\lambda)-1)P_{k}(x,t), \qquad c_{k}(\lambda)= \frac{\lambda-\lambda_{k}^{+}}{\lambda-\lambda^{-}_{k}}. \label{The Dressing Method: The Singular SolutionsLocal2b}
\end{equation}
Here  $u_{k}(x,t,\lambda)$ is the dressing factor, $P_{k}(x,t)$ is a projector of rank $1$ has the form:
\begin{align}\label{The Dressing Method: The Singular SolutionsLocal3}
P_{k}(x,t) = \frac{|n_k(x,t)\rangle \langle m_k(x,t)|}{\langle m_k(x,t)|n_k(x,t)\rangle},
\end{align}
where $|n_k(x,t)\rangle =\chi^{+}_{0}(x,t,\lambda^{+}_{1})|n_{0,1}\rangle$ is a column vector and $\langle m_{1}(x,t)|=  \langle m_{0,1}| \hat{\chi}^{-}_{0}(x,t,\lambda^{-}_{1})$ is a row vector and both of them are eigenvector. The projectors $P_{k}(x,t)$  automatically satisfy the condition $P_{k}^{2}(x,t)=P_{k}(x,t)$.
The functions $\eta^{\pm}(x,t,\lambda)$  satisfies the linear system:
\begin{equation}\label{The Zakharov-Shabat Dressing Method Local12}
\left({\rm i}\frac{\rd \eta^\pm}{\rd x}+Q(x,t)\eta^\pm(x,t,\lambda)- \lambda\left[\rJ, \eta^\pm(x,t,\lambda)\right]\right)=0.
\end{equation}
where $\eta^{\pm}(x,t,\lambda)$ are related to the ZS system with an unknown potential $Q(x,t)$, which can be found later, while $\eta_{0}^{\pm}(x,t,\lambda)$ is related to the ZS system with a known potential $Q_{0}(x,t)$:
\begin{equation}\label{The Dressing Method: The Singular SolutionsLocal11}
{\rm i}\frac{\rd \eta_{0}^\pm}{\rd x}+Q_{0}(x,t)\eta_{0}^\pm(x,t,\lambda)-(\lambda)\left[\rJ, \eta_{0}^\pm(x,t,\lambda)\right]=0.
\end{equation}
Now, from \eqref{The Dressing Method: The Singular SolutionsLocal2a} and \eqref{The Dressing Method: The Singular SolutionsLocal11}, we find that the dressing factor $u_k(x,t,\lambda)$ satisfies the following equation:
\begin{equation}\label{The Dressing Method: The Singular SolutionsLocal12}
{\rm i}\frac{\rd u_k}{\rd x}+Q(x,t)u_k(x,t,\lambda)-u_k(x,t,\lambda)Q_{0}(x,t)- \lambda \left[\rJ, u_k(x,t,\lambda)\right]=0,
\end{equation}
since the anzats for the dressing factor $u_k(x,t,\lambda)$ in \eqref{The Dressing Method: The Singular SolutionsLocal2b} and \eqref{The Dressing Method: The Singular SolutionsLocal12} are compatible with respect to $\lambda$, then there are two conditions can be applied to the left-hand side of \eqref{The Dressing Method: The Singular SolutionsLocal12} which are the limit for $\lambda\rightarrow\infty$ and the residue at $\lambda=\lambda^{-}_{k}$ and both of them are vanish. The first condition leads to the potential:
\begin{align}\label{The Dressing Method: The Singular SolutionsLocal13}
Q(x,t)-Q_{0}(x,t)=-(\lambda^{+}_{k}-\lambda^{-}_{k})[\rJ,P_k(x,t)],
\end{align}
and the second condition gives the following nonlinear equation for $P_k(x,t)$:
\begin{align}
{\rm i}\frac{\rd P_{k}}{\rd x}& + Q_{0}(x,t)P_{k}(x,t)-P_{k}(x,t)Q_{0}(x,t)-\lambda^{+}_{k}\rJ P_{k}(x,t)+ \lambda^{-}_{k} P_{k}(x,t) \rJ
\\
&+ (\lambda^{+}_{k}-\lambda^{-}_{k})P_{k}(x,t) \rJ P_{k}(x) =0. \label{The Dressing Method: The Singular SolutionsLocal14}
\end{align}
In addition, we have  the normalization condition $\lim_{\lambda\to \infty} u_k (x,t,\lambda)= \mathds{1}$.

In the case of ${\bf BD.I}$ symmetric spaces the dressing factor $u_k(x,t,\lambda)$ can be written in the form:
\begin{align}
u_{k} (x,t,\lambda)= & \mathds{1} + (c_{k}(\lambda)-1)P_{k}(x,t)+  \left(\frac{1}{c_{k}(\lambda)}-1\right)\overline{P}_{k}(x,t),  \qquad \overline{P}_{k}=S_{0}P^{T}_{k}S_{0}^{-1}, \label{One Soliton Solution:so(5)b}
\end{align}
where $P_{k}(x,t)$ and $\overline{P}_{k}(x,t)$ are mutually orthogonal projectors with rank $1$ \cite{vgrn}. In a similar fashion, one can write the dressed potential $Q(x,t)$ as:
\begin{align}\label{One Soliton Solution:so(5)4}
Q(x,t)= Q_{0}(x,t)-(\lambda^{+}_{k}-\lambda^{-}_{k}) [\rJ,P_k(x,t)-\overline{P}_{k}(x,t)].
\end{align}

\subsection{One Soliton Solutions}\label{sec:4.1}

\begin{example} If we take ${\frak g}\simeq sl(3,{\Bbb C})$ and a dressing factor in the form  \eqref{The Dressing Method: The Singular SolutionsLocal2b} satisfying the nonlocal reduction conditions \eqref{The Local Involutions2m1}, this will imply the following involution on the dressing factor:
\begin{align}\label{The Nonlocal Involution Case:g=sl(3)}
B \ u_{1}(-x,t,-\lambda^{\ast})^{\dagger}\ B^{-1}=& \ u^{-1}_{1}(x,t,\lambda),
\end{align}
where $B$ is constant block-diagonal matrix. As a result, the projector $P_1$ must satisfy:
\begin{align}\label{The Nonlocal Involution Case:g=sl(3)1}
P_{1}(x,t)= & BP_{1}^{\dagger}(-x,t)B^{-1},\qquad (-\lambda^{\pm})^{\ast}=\lambda^{\mp}.
\end{align}
That means the projector $P_{1}(x,t)$  and $c_{1}(\lambda)$ become:
\begin{align}\label{The Nonlocal Involution Case:g=sl(3)2}
P_{1}(x,t)= & \frac{|n_1(x,t)\rangle \langle n_1^{\ast}(-x,t)|B}{\langle n_1^{\ast}(-x,t)|B | n_1(x,t)\rangle}, \qquad \langle m_1(x,t)|=(B | n_1(-x,t)\rangle)^{\dagger},
\\
c_{1}(\lambda)= & \frac{\lambda-\lambda_{1}^{+}}{\lambda+(\lambda^{+}_{1})^{\ast}}.
\end{align}
So, from \eqref{The Dressing Method: The Singular SolutionsLocal13}, the components of the one-soliton solution can be written as:
\begin{subequations}\label{The Nonlocal Involution Case:g=sl(3)6}
\begin{align}\label{The Nonlocal Involution Case:g=sl(3)6a}
q_{1j}(x,t)= &-2(\lambda^{+}_{1}+(\lambda^{+}_{1})^{\ast})  \  \frac {n_{0,1}^{1} (n_{0,1}^{j})^{\ast}  \re^{-\ri( \tilde{M}_{1}^{+}(x,t)+ (\tilde{M}_{1}^{+})^{*}(-x,t))}}{2 R_{0,1} \rm {cosh} (2\nu_{1}x+2 \tilde{\mu_{1}}t+\xi_{0,1})}, \qquad j=2,3, \nonumber
\\
q_{j1}(x,t)= & \ -2(\lambda^{+}_{1}+(\lambda^{+}_{1})^{\ast})  \ \frac {n_{0,1}^{j} (n_{0,1}^{1})^{\ast}  \re^{\ri(\tilde{M}_{1}^{+}(x,t)+ (\tilde{M}_{1}^{+})^{*} (-x,t))}}{2 R_{0,1} \rm {cosh} (2\nu_{1}x+2\tilde{\mu_{1}}t+ \xi_{0,1})}, \quad \qquad j=2,3.
\end{align}
Here,
\begin{align}
\tilde{M}_{1}^{\pm}(x,t)= & \lambda_{1}^{\pm}x + 2 \ \rm f_{0, 1}^{\pm} t, \qquad (\tilde{M}_{1}^{+}(-x,t))^{*} = (-\lambda_{1}^{+})^{*}(-x) + 2 \ \rm (f_{0}(-\lambda_{1}^{+})^{\ast}) t, \label{The Nonlocal Involution Case:g=sl(3)6b}
\\
\nu_{1}= & \rm i ((-\lambda^{+}_{1})^{*}-\lambda^{+}_{1})/2, \qquad  \tilde{\mu_{1}}= \rm i (\rm f(-\lambda^{+}_{1})^{*}-\rm f(\lambda^{+}_{1})), \label{The Nonlocal Involution Case:g=sl(3)6c}
\\
R_{0,1}= & \sqrt{(n_{0,1}^{1})^{\ast}n_{0,1}^1 ( (n_{0,1}^{2})^{\ast}n_{0,1}^2  +  (n_{0,1}^{3})^{\ast}n_{0,1}^3)}, \quad  \xi_{0,1}=\frac{1}{2}\rm {ln} \frac{ (n_{0,1}^{1})^{\ast}n_{0,1}^1}{(n_{0,1}^{2})^{\ast}n_{0,1}^2+  (n_{0,1}^{3})^{\ast}n_{0,1}^3}. \label{The Nonlocal Involution Case:g=sl(3)6d}
\end{align}
\end{subequations}
\end{example}

\begin{example} If we take again ${\frak g}\simeq sl(3,{\Bbb C})$ and the involution automorphism to be the Weyl reflection with respect to the second simple root ${\rm e}_2-{\rm e}_3$ of $sl(3,{\Bbb C})$ and impose a reduction  of type (B) in \eqref{The Local Involutions2m1}. This will correspond to another block matrix $B = \left(
\begin{array}{cc}
\mathbf{B_+}  &0 \\
0& \mathbf{B_-} \\
\end{array}\right)$ where $ \mathbf{B_-}$ is a block off-diagonal matrix. The blocks can be written as
\begin{align}\label{another nonlocal involution}
\mathbf{B_+}=\mp1, \qquad \text{and} \qquad  \mathbf{B_-}= \left(\begin{array}{cc}
0&\pm1 \\
\pm1&0\\
\end{array}\right).
\end{align}
The reductions conditions for the matrix blocks can be written as:
\begin{align}\label{another nonlocal involution1}
\mbox{(C)} &&\qquad
\mathbf{q}^{-}(x,t)=-\mathbf{B_-}(\mathbf{q}^{+}(-x,t))^{\dag}(\mathbf{B_+})^{-1},\quad \mathbf{q}^{+}(x,t)=-\mathbf{B_+}(\mathbf{q}^{-}(-x,t))^{\dag}(\mathbf{B_-})^{-1}.
\end{align}
Now, if we apply the involution of type (C) with using \eqref{another nonlocal involution}, then the standard Manakov VNLS equation reproduced as:

\begin{equation}\label{another nonlocal involution2}
\begin{array}{cc}
-{\rm i}q_{1,t}+ q_{1,xx}+2 \ (q_{1}(x,t) \ q_{2}^{\ast}(-x,t)+ q_{2}(x,t) \ q_{1}^{\ast}(-x,t)) \ q_{1}(x,t)=0,
\vspace{0.1in}\\
-{\rm i}q_{2,t}+ q_{2,xx}+2 \ (q_{1}(x,t) \ q_{2}^{\ast}(-x,t)+ q_{2}(x,t) \ q_{1}^{\ast}(-x,t)) \  q_{2}(x,t)=0.
\end{array}
\end{equation}
Also, we can write one-soliton solution of the Manakov VNLS equation for this type of involution as the dressing factor \eqref{The Dressing Method: The Singular SolutionsLocal2b} is satisfied by the involution \eqref{The Nonlocal Involution Case:g=sl(3)} with \eqref{The Nonlocal Involution Case:g=sl(3)1}. So, from \eqref{The Dressing Method: The Singular SolutionsLocal13}, the components of the one-soliton solution can be written as:
\begin{subequations}\label{another nonlocal involution3}
\begin{align}\label{another nonlocal involution3a}
q_{12}(x,t)= &-2(\lambda^{+}_{1}+(\lambda^{+}_{1})^{\ast}) \ n_{0,1}^{1} \re^{-\ri \tilde{M}_{1}^{+}(x,t)} R_{1}^{-1}(x,t)(n_{0,1}^{3})^{\ast}  \re^{-\ri\tilde{M}_{1}^{+,\ast}(-x,t)}, \nonumber
\\
q_{13}(x,t)= &-2(\lambda^{+}_{1}+(\lambda^{+}_{1})^{\ast}) \ n_{0,1}^{1} \re^{-\ri \tilde{M}_{1}^{+}(x,t)} R_{1}^{-1}(x,t)(n_{0,1}^{2})^{\ast}  \re^{-\ri\tilde{M}_{1}^{+,\ast}(-x,t)}, \nonumber
\\
q_{21}(x,t)= & \ -2(\lambda^{+}_{1}+(\lambda^{+}_{1})^{\ast})  \ n_{0,1}^{2} \re^{\ri \tilde{M}_{1}^{+}(x,t)} R_{1}^{-1}(x,t) (n_{0,1}^{1})^{\ast} \re^{\ri \tilde{M}_{1}^{+,\ast}(-x,t))}, \nonumber
\\
q_{31}(x,t)= & \ -2(\lambda^{+}_{1}+(\lambda^{+}_{1})^{\ast})  \ n_{0,1}^{3} \re^{\ri \tilde{M}_{1}^{+}(x,t)} R_{1}^{-1}(x,t) (n_{0,1}^{1})^{\ast} \re^{\ri \tilde{M}_{1}^{+,\ast}(-x,t))}.
\end{align}
Here,
\begin{align}
\tilde{M}_{1}^{\pm}(x,t)= & \lambda_{1}^{\pm}x + 2 \ \rm f_{0, 1}^{\pm} t, \qquad (\tilde{M}_{1}^{+}(-x,t))^{*} = (-\lambda_{1}^{+})^{*}(-x) + 2 \ \rm (f_{0}(-\lambda_{1}^{+})^{\ast}) t, \label{another nonlocal involution3b}
\\
R_{1}(x,t)= &2 R_{0,1} \rm {cosh} (2\nu_{1}x+2\tilde{\mu_{1}}t+ \xi_{0,1}), \label{another nonlocal involution3c}
\\
\nu_{1}= &\rm i ((-\lambda^{+}_{1})^{*}-\lambda^{+}_{1})/2, \qquad  \tilde{\mu_{1}}= \rm i (\rm f(-\lambda^{+}_{1})^{*}-\rm f(\lambda^{+}_{1})), \label{another nonlocal involution3c1}
\\
R_{0,1}= & \sqrt{(n_{0,1}^{1})^{\ast}n_{0,1}^1 ( (n_{0,1}^{3})^{\ast}n_{0,1}^2  +  (n_{0,1}^{2})^{\ast}n_{0,1}^3)}, \qquad  \xi_{0,1}=\frac{1}{2}\rm {ln} \frac{ (n_{0,1}^{1})^{\ast}n_{0,1}^1}{(n_{0,1}^{3})^{\ast}n_{0,1}^2+  (n_{0,1}^{2})^{\ast}n_{0,1}^3}. \label{another nonlocal involution3d}
\end{align}
\end{subequations}
\end{example}

\begin{example} If we take ${\frak g}\simeq so(5,{\Bbb C})$, then  the dressing factor will take the form \eqref{One Soliton Solution:so(5)b}. If we impose  the nonlocal reduction conditions \eqref{The Local Involutions2m1}:
\begin{align}\label{The nonlocal Involution Case:g=so(5)}
B \ u_{1}(-x,t,-\lambda^{\ast})^{\dagger}\ B^{-1}=& \ u^{-1}_{1}(x,t,\lambda),
\end{align}
this will lead to the following reduction condition:
\begin{align}\label{The nonlocal Involution Case:g=so(5)1}
P_{1}(x,t)= & BP_{1}^{\dagger}(-x,t)B^{-1},\qquad (-\lambda^{\pm})^{\ast}=\lambda^{\mp}.
\end{align}
That means the projector $P_{1}(x,t)$ and $c_{1}(\lambda)$ become:
\begin{align}
P_{1}(x,t)= & \frac{|n_1(x,t)\rangle \langle n_1^{\ast}(-x,t)|B}{\langle n_1^{\ast}(-x,t)|B | n_1(x,t)\rangle}, \qquad \langle m_1(x,t)|=(B | n_1(-x,t)\rangle)^{\dagger},\label{The nonlocal Involution Case:g=so(5)2a}
\\
c_{1}(\lambda)= & \frac{\lambda-\lambda_{1}^{+}}{\lambda+(\lambda^{+}_{1})^{\ast}}.\label{The nonlocal Involution Case:g=so(5)2b}
\end{align}
Thus, taking the trivial solution $Q_{0}(x,t)=0$,  the components of one soliton solution can be written as:
\begin{subequations}\label{The nonlocal Involution Case:g=so(5)4The nonlocal Involution Case:g=so(5)4}
\begin{align}
q_{12}^{+}(x,t)= & q_{45}^{+}(x,t)= -(\lambda^{+}_{1}+(\lambda^{+}_{1})^{\ast})R_1^{-1}(x,t)(n_{0,1}^{1} (n_{0,1}^{2})^{\ast} \re^{-\ri M_{1}^{+}(x,t)} -n_{0,1}^{4}(n_{0,1}^{5})^{\ast} \re^{-\ri (M_{1}^{+}(-x,t))^{\ast}}), \nonumber
\\
q_{13}^{+}(x,t)= &- q_{35}^{+}(x,t)= -(\lambda^{+}_{1}+(\lambda^{+}_{1})^{\ast})R_1^{-1}(x,t)(n_{0,1}^{1} (n_{0,1}^{3})^{\ast}  \re^{-\ri M_{1}^{+}(x,t)} +n_{0,1}^{3}(n_{0,1}^{5})^{\ast} \re^{-\ri (M_{1}^{+}(-x,t))^{\ast}}), \nonumber
\\
q_{14}^{+}(x,t)= & q_{25}^{+}(x,t)= -(\lambda^{+}_{1}+(\lambda^{+}_{1})^{\ast})R_1^{-1}(x,t)(n_{0,1}^{1} (n_{0,1}^{4})^{\ast}  \re^{-\ri M_{1}^{+}(x,t)} -n_{0,1}^{2} (n_{0,1}^{5})^{\ast} \re^{-\ri (M_{1}^{+}(-x,t))^{\ast}}), \nonumber
\\
q_{21}^{-}(x,t)= & q_{54}^{-}(x,t)= -(\lambda^{+}_{1}+(\lambda^{+}_{1})^{\ast})R_1^{-1}(x,t)(n_{0,1}^{2} (n_{0,1}^{1})^{\ast}  \re^{\ri (M_{1}^{-}(-x,t))^{\ast}}-n_{0,1}^{5}(n_{0,1}^{4})^{\ast} \re^{\ri M_{1}^{+}(x,t)}), \nonumber
\\
q_{31}^{-}(x,t)= & - q_{53}^{-}(x,t)= -(\lambda^{+}_{1}+(\lambda^{+}_{1})^{\ast})R_1^{-1}(x,t)(n_{0,1}^{3} (n_{0,1}^{1})^{\ast}  \re^{\ri (M_{1}^{-}(-x,t))^{\ast}} +n_{0,1}^{5}(n_{0,1}^{3})^{\ast} \re^{\ri M_{1}^{+}(x,t)}), \nonumber
\\
q_{41}^{-}(x,t)= & q_{52}^{-}(x,t)= -(\lambda^{+}_{1}+(\lambda^{+}_{1})^{\ast})R_1^{-1}(x,t)(n_{0,1}^{4} (n_{0,1}^{1})^{\ast} \re^{\ri (M_{1}^{+}(-x,t))^{\ast}}-n_{0,1}^{5}(n_{01}^{2})^{\ast} \re^{\ri M_{1}^{+}(x,t)}).\label{The nonlocal Involution Case:g=so(5)4a}
\end{align}
Here,
\begin{align}
(M_{1}^{+}(-x,t))^{\ast}= &(-\lambda_{1}^{+})^{\ast}(-x) + 2 \ (\rm f(-\lambda^{+}_{1})^{*}) t, \quad R_{1}(x,t)=2 R_{0,1} \rm {cosh} (2\nu_{1}(x)+2\tilde{\mu_{1}}t+ \xi_{0,1})+S, \label{The nonlocal Involution Case:g=so(5)4b}
\\
\nu_{1}= & \rm i ((-\lambda^{+}_{1})^{*}-\lambda^{+}_{1})/2, \qquad  \tilde{\mu_{1}}= \rm i (\rm f(-\lambda^{+}_{1})^{*}-\rm f(\lambda^{+}_{1})),\label{The nonlocal Involution Case:g=so(5)4c}
\\
S= &(n_{0,1}^{2})^{*} n_{0,1}^{2}+ (n_{0,1}^{3})^{*} n_{0,1}^{3}+ (n_{0,1}^{4})^{*} n_{0,1}^{4},\label{The nonlocal Involution Case:g=so(5)4d}
\\
R_{0,1}= & \sqrt{  (n_{0,1}^1)^{*} n_{0,1}^1  +  (n_{0,1}^5)^{*} n_{0,1}^5 }, \qquad  \xi_{0,1}=\frac{1}{2}\rm {ln} \frac{ (n_{0,1}^1)^{*} n_{0,1}^1 }{  (n_{0,1}^5)^{*} n_{0,1}^5 }. \label{The nonlocal Involution Case:g=so(5)4e}
\end{align}
\end{subequations}
\end{example}
\begin{example} If we take again ${\frak g}\simeq so(5,{\Bbb C})$ and another  $\mathbb{Z}_2$ reduction of type (B) \eqref{The Local Involutions2m1} with a  block matrix $B$ containing an off-diagonal block $\mathbf{B_\pm}$: $\mathbf{B_+}=\mathbf{B_-}=\mp1$ and $\mathbf{B_\pm}= \text{off-diag}(\pm1,\ldots,\pm1)$.
In this case, the reductions conditions for the matrix blocks can be written as:
\begin{align}\label{another nonlocal involution1a}
\mathbf{q}^{-}(x,t)=-\mathbf{B_-}(\mathbf{q}^{+}(-x,t))^{\dag}(\mathbf{B_+})^{-1},\quad \mathbf{q}^{+}(x,t)=-\mathbf{B_+}(\mathbf{q}^{-}(-x,t))^{\dag}(\mathbf{B_-})^{-1}.
\end{align}
Now, recall example A from Section 2, if we apply the involution of type (B) with using $\mathbf{B_+}=\mathbf{B_-}=\mp1$ and $\mathbf{B_\pm}= \text{off-diag}(\pm1,\ldots,\pm1)$, then the 3-component NLS system in \eqref{Some Algebraic Properties for Cartan Weyl Basis7} can be rewritten as:
\begin{align}\label{another nonlocal involutionso(5)}
\ri(q_{t}^{+})_{12}&+(q_{xx}^{+})_{12}+
2\left(q_{12}(x,t)q_{14}^{\ast}(-x,t)+2 q_{13}(x,t)q_{13}^{\ast}(-x,t)\right) q_{12} +2 q^{+,\ast}_{12}(-x,t)(q^{+}_{13})^{2}=0 ,\nonumber
\\
\ri(q_{t}^{+})_{13}&+(q_{xx}^{+})_{13}+
2\left(q_{12}(x,t)q_{14}^{\ast}(-x,t)+ q_{13}(x,t)q_{13}^{\ast}(-x,t)+ q_{14}(x,t)q_{12}^{\ast}(-x,t) \right) q_{13}\nonumber
\\
&+2 q^{+,\ast}_{13}(-x,t)q^{+}_{14}q^{+}_{12}=0 ,\nonumber
\\
\ri(q_{t}^{+})_{14}&+(q_{xx}^{+})_{14}+ 2\left(q_{14}(x,t)q_{12}^{\ast}(-x,t)+2 q_{13}(x,t)q_{13}^{\ast}(-x,t) \right) q_{14} +2 q^{+,\ast}_{14}(-x,t)(q^{+}_{13})^{2}=0.
\end{align}
Also, we can write one-soliton solution of this system for this type of involution as the dressing factor \eqref{One Soliton Solution:so(5)b} is satisfied by the involution \eqref{The nonlocal Involution Case:g=so(5)} with \eqref{The nonlocal Involution Case:g=so(5)1}. So, from \eqref{One Soliton Solution:so(5)4}, the components of the one-soliton solution can be written as:
\begin{subequations}\label{The nonlocal Involution Case:g=so(5)4x}
\begin{align}
q_{12}^{+}(x,t)= & q_{45}^{+}(x,t)= -(\lambda^{+}_{1}+(\lambda^{+}_{1})^{\ast})R_1^{-1}(x,t)(n_{0,1}^{1} (n_{0,1}^{4})^{\ast} \re^{-\ri M_{1}^{+}(x,t)} -n_{0,1}^{4}(n_{0,1}^{5})^{\ast} \re^{-\ri (M_{1}^{+}(-x,t))^{\ast}}), \nonumber
\\
q_{13}^{+}(x,t)= &- q_{35}^{+}(x,t)= -(\lambda^{+}_{1}+(\lambda^{+}_{1})^{\ast})R_1^{-1}(x,t)(n_{0,1}^{1} (n_{0,1}^{3})^{\ast}  \re^{-\ri M_{1}^{+}(x,t)} +n_{0,1}^{3}(n_{0,1}^{5})^{\ast} \re^{-\ri (M_{1}^{+}(-x,t))^{\ast}}), \nonumber
\\
q_{14}^{+}(x,t)= & q_{25}^{+}(x,t)= -(\lambda^{+}_{1}+(\lambda^{+}_{1})^{\ast})R_1^{-1}(x,t)(n_{0,1}^{1} (n_{0,1}^{2})^{\ast}  \re^{-\ri M_{1}^{+}(x,t)} -n_{0,1}^{2} (n_{0,1}^{5})^{\ast} \re^{-\ri (M_{1}^{+}(-x,t))^{\ast}}), \nonumber
\\
q_{21}^{-}(x,t)= & q_{54}^{-}(x,t)= -(\lambda^{+}_{1}+(\lambda^{+}_{1})^{\ast})R_1^{-1}(x,t)(n_{0,1}^{2} (n_{0,1}^{1})^{\ast}  \re^{\ri (M_{1}^{-}(-x,t))^{\ast}}-n_{0,1}^{5}(n_{0,1}^{2})^{\ast} \re^{\ri M_{1}^{+}(x,t)}), \nonumber
\\
q_{31}^{-}(x,t)= & - q_{53}^{-}(x,t)= -(\lambda^{+}_{1}+(\lambda^{+}_{1})^{\ast})R_1^{-1}(x,t)(n_{0,1}^{3} (n_{0,1}^{1})^{\ast}  \re^{\ri (M_{1}^{-}(-x,t))^{\ast}} +n_{0,1}^{5}(n_{0,1}^{3})^{\ast} \re^{\ri M_{1}^{+}(x,t)}), \nonumber
\\
q_{41}^{-}(x,t)= & q_{52}^{-}(x,t)= -(\lambda^{+}_{1}+(\lambda^{+}_{1})^{\ast})R_1^{-1}(x,t)(n_{0,1}^{4} (n_{0,1}^{1})^{\ast} \re^{\ri (M_{1}^{+}(-x,t))^{\ast}}-n_{0,1}^{5}(n_{01}^{4})^{\ast} \re^{\ri M_{1}^{+}(x,t)}).\label{The nonlocal Involution Case:g=so(5)4ax}
\end{align}
Here,
\begin{align}
(M_{1}^{+}(-x,t))^{\ast}= &(-\lambda_{1}^{+})^{\ast}(-x) + 2 \ (\rm f(-\lambda^{+}_{1})^{*}) t, \quad R_{1}(x,t)=2 R_{0,1} \rm {cosh} (2\nu_{1}(x)+2\tilde{\mu_{1}}t+ \xi_{0,1})+S, \label{The nonlocal Involution Case:g=so(5)4bx}
\\
\nu_{1}= & \rm i ((-\lambda^{+}_{1})^{*}-\lambda^{+}_{1})/2, \qquad  \tilde{\mu_{1}}= \rm i (\rm f(-\lambda^{+}_{1})^{*}-\rm f(\lambda^{+}_{1})),\label{The nonlocal Involution Case:g=so(5)4cx}
\\
S= &(n_{0,1}^{4})^{*} n_{0,1}^{2}+ (n_{0,1}^{3})^{*} n_{0,1}^{3}+ (n_{0,1}^{2})^{*} n_{0,1}^{4},\label{The nonlocal Involution Case:g=so(5)4dx}
\\
R_{0,1}= & \sqrt{  (n_{0,1}^1)^{*} n_{0,1}^1  +  (n_{0,1}^5)^{*} n_{0,1}^5 }, \qquad  \xi_{0,1}=\frac{1}{2}\rm {ln} \frac{ (n_{0,1}^1)^{*} n_{0,1}^1 }{  (n_{0,1}^5)^{*} n_{0,1}^5 }. \label{The nonlocal Involution Case:g=so(5)4ex}
\end{align}
\end{subequations}
\end{example}

\subsection{Two Soliton Solution}\label{sec:4.2}

\begin{example} For ${\frak g}\simeq sl(3,{\Bbb C})$, in order to get the 2-soliton solution, the dressing factor c \eqref{The Dressing Method: The Singular SolutionsLocal2} can be rewritten as:
\begin{subequations}\label{involution of two Soliton Solution}
\begin{align}
\eta^{+}(x,t,\lambda)= & u_{1,2} (x,t,\lambda)\eta^{+}_{0}(x,t,\lambda) (\omega ^{\pm}_{1,2}(\lambda))^{-1} , \label{involution of two Soliton Solutiona}
\\
u_{1,2}(x,t,\lambda) = & \mathds{1} + (c_{1}(\lambda)-1)P_{1}(x,t)+ (c_{2}(\lambda)-1)P_{2}(x,t).\label{involution of two Soliton Solutionb}
\end{align}
\end{subequations}
Thus, the singular solution (two soliton solution) with singularities located at $\lambda_{1}^{\pm}$ and $\lambda_{2}^{\pm}$ can be obtained as follows:
\begin{align}\label{involution of two Soliton Solution2}
Q(x,t)=& Q_{0}(x,t)-\sum^{2}_{k=1}(\lambda^{+}_{k}-\lambda^{-}_{k})[\rJ,P_k(x,t)].
\end{align}
Assuming a canonical reduction of type (B) \eqref{The Local Involutions2m1} with ${\bf B}={\Bbb I}$, from \eqref{involution of two Soliton Solution2} the components of the two-soliton solution can be written as:
\begin{subequations}\label{involution of two Soliton Solution6}
\begin{align}
q_{1j}(x,t)= &-2\sum^{2}_{k=1}(\lambda^{+}_{k}+(\lambda^{+}_{k})^{\ast})  \  \frac {n_{0,k}^{k} (n_{0,k}^{j})^{\ast}  \re^{-\ri( \tilde{M}_{k}^{+}(x,t)+ (\tilde{M}_{k}^{+})^{*}(-x,t))}}{2 R_{0,k} \rm {cosh} (2\nu_{k}x+2 \tilde{\mu_{k}}t+\xi_{0,k})}, \qquad j=2,3, \nonumber
\\
q_{j1}(x,t)= & \- 2\sum^{2}_{k=1}(\lambda^{+}_{k}+(\lambda^{+}_{k})^{\ast})  \ \frac {n_{0,k}^{j} (n_{0,k}^{1})^{\ast}  \re^{\ri(\tilde{M}_{k}^{+}(x,t)+ (\tilde{M}_{k}^{+})^{*} (-x,t))}}{2 R_{0,k} \rm {cosh} (2\nu_{k}x+2\tilde{\mu_{k}}t+ \xi_{0,k})}, \quad \qquad j=2,3. \label{involution of two Soliton Solution6a}
\end{align}
Here,
\begin{align}
\tilde{M}_{k}^{\pm}(x,t)= & \lambda_{k}^{\pm}x + 2 \ \rm f_{0, k}^{\pm} t, \qquad (\tilde{M}_{k}^{+}(-x,t))^{*} = (-\lambda_{k}^{+})^{*}(-x) + 2 \ (\rm f(-\lambda^{+}_{1})^{*}) t, \label{involution of two Soliton Solution6b}
\\
\nu_{k}= & \rm i ((-\lambda^{+}_{k})^{*}-\lambda^{+}_{k})/2, \qquad  \tilde{\mu_{k}}= \rm i (\rm f(-\lambda^{+}_{k})^{*}-\rm f(\lambda^{+}_{k})), \label{involution of two Soliton Solution6c}
\\
R_{0,k}= & \sqrt{(n_{0,k}^{1})^{\ast}n_{0,k}^1 ( (n_{0,k}^{2})^{\ast}n_{0,k}^2  +  (n_{0,k}^{3})^{\ast}n_{0,k}^3)}, \quad  \xi_{0,k}=\frac{1}{2}\rm {ln} \frac{ (n_{0,k}^{1})^{\ast}n_{0,k}^1}{(n_{0,k}^{2})^{\ast}n_{0,k}^2+  (n_{0,k}^{3})^{\ast}n_{0,k}^3}. \label{involution of two Soliton Solution6d}
\end{align}
\end{subequations}

\end{example}

\begin{remark}
Alternatively, in order to obtain 2-soliton solution, one can apply dressing method again on the 1-soliton solution used as a seed solution:
\begin{align}\label{involution of two Soliton Solution1}
\eta^{+}(x,t,\lambda)= & u_{2}(x,t,\lambda)\eta^{+}_{1}(x,t,\lambda)(\omega ^{\pm}_{2}(\lambda))^{-1}, \qquad  u_{2} (x,t,\lambda)=\mathds{1} + (c_{2}(\lambda)-1)P_{2}(x,t).
\end{align}
\end{remark}

\begin{example} In the case of the involution of  Example 2,  the 2-soliton dressing factor \eqref{involution of two Soliton Solutiona} is automatically compatible with the involution. Therefore, from \eqref{involution of two Soliton Solution2}, the components of the two-soliton solution can be written as:
\begin{subequations}\label{two another nonlocal involution sl(3)}
\begin{align}
q_{12}(x,t)= &-2\sum^{2}_{k=1}(\lambda^{+}_{k}+(\lambda^{+}_{k})^{\ast}) \ n_{0,k}^{1} \re^{-\ri \tilde{M}_{k}^{+}(x,t)} R_{k}^{-1}(x,t)(n_{0,k}^{3})^{\ast}  \re^{-\ri\tilde{M}_{k}^{+,\ast}(-x,t)}, \nonumber
\\
q_{13}(x,t)= &-2\sum^{2}_{k=1}(\lambda^{+}_{k}+(\lambda^{+}_{k})^{\ast}) \ n_{0,k}^{1} \re^{-\ri \tilde{M}_{k}^{+}(x,t)} R_{k}^{-1}(x,t)(n_{0,k}^{2})^{\ast}  \re^{-\ri\tilde{M}_{k}^{+,\ast}(-x,t)}, \nonumber
\\
q_{21}(x,t)= & \ -2\sum^{2}_{k=1}(\lambda^{+}_{k}+(\lambda^{+}_{k})^{\ast})  \ n_{0,k}^{2} \re^{\ri \tilde{M}_{k}^{+}(x,t)} R_{k}^{-1}(x,t) (n_{0,k}^{1})^{\ast} \re^{\ri \tilde{M}_{k}^{+,\ast}(-x,t))}, \nonumber
\\
q_{31}(x,t)= & \ -2\sum^{2}_{k=1}(\lambda^{+}_{k}+(\lambda^{+}_{k})^{\ast})  \ n_{0,k}^{3} \re^{\ri \tilde{M}_{k}^{+}(x,t)} R_{k}^{-1}(x,t) (n_{0,k}^{1})^{\ast} \re^{\ri \tilde{M}_{k}^{+,\ast}(-x,t))}. \label{two another nonlocal involution sl(3)a}
\end{align}
Here,
\begin{align}
\tilde{M}_{k}^{\pm}(x,t)= & \lambda_{k}^{\pm}x + 2 \ \rm f_{0, k}^{\pm} t, \qquad (\tilde{M}_{k}^{+}(-x,t))^{*} = (-\lambda_{k}^{+})^{*}(-x) + 2 \ (\rm f(-\lambda^{+}_{1})^{*}) t, \label{two another nonlocal involution sl(3)b}
\\
R_{1}(x,t)= &2 R_{0,1} \rm {cosh} (2\nu_{1}x+2\tilde{\mu_{1}}t+ \xi_{0,1}),\label{two another nonlocal involution sl(3)c}
\\
\nu_{k}= & \rm i ((-\lambda^{+}_{k})^{*}-\lambda^{+}_{k})/2, \qquad  \tilde{\mu_{k}}= \rm i (\rm f(-\lambda^{+}_{k})^{*}-\rm f(\lambda^{+}_{k})), \label{two another nonlocal involution sl(3)c1}
\\
R_{0,k}= & \sqrt{(n_{0,k}^{1})^{\ast}n_{0,k}^1 ( (n_{0,k}^{3})^{\ast}n_{0,k}^2  +  (n_{0,k}^{2})^{\ast}n_{0,k}^3)}, \qquad  \xi_{0,k}=\frac{1}{2}\rm {ln} \frac{ (n_{0,k}^{1})^{\ast}n_{0,k}^1}{(n_{0,k}^{3})^{\ast}n_{0,k}^2+  (n_{0,k}^{2})^{\ast}n_{0,k}^3}. \label{two another nonlocal involution sl(3)d}
\end{align}
\end{subequations}

\end{example}

\begin{example}If we take ${\frak g}\simeq so(5,{\Bbb C})$, then the 2-soliton dressing factors $u(x,t,\lambda)$ can be considered with two more poles in the following form:
\begin{align}\label{Two Soliton Solution:so(5)}
u_{1,2} (x,t,\lambda)= & \mathds{1} + (c_{1}(\lambda)-1)P_{1}(x,t)+  \left(\frac{1}{c_{1}(\lambda)}-1\right)\overline{P}_{1}(x,t),\nonumber
\\
& + (c_{2}(\lambda)-1)P_{2}(x,t)+  \left(\frac{1}{c_{2}(\lambda)}-1\right)\overline{P}_{2}(x,t),
\\
c_{k}(\lambda)=  \frac{\lambda-\lambda_{k}^{+}}{\lambda-\lambda^{-}_{k}} & , \quad  P_{k}(x,t) = \frac{|n_k(x,t)\rangle \langle m_k(x,t)|}{\langle m_k(x,t)|n_k(x,t)\rangle}, \quad \overline{P}_{k}=S_{0}P^{T}_{k}S_{0}^{-1}, \quad k=1,2.\label{Two Soliton Solution:so(5)a}
\end{align}
Thus, the new potential $Q(x,t)$ corresponding to a given trivial solution $Q_{0}(x,t)=0$ can be obtained by the following form:
\begin{align}\label{Two Soliton Solution:so(5)1}
Q(x,t)= -\sum_{k=1}^{2}(\lambda^{+}_{k}-\lambda^{-}_{k}) [\rJ,P_k(x,t)-\overline{P}_{k}(x,t)].
\end{align}
The components of new potential $Q(x,t)$ can be obtained by:
\begin{subequations}\label{Two Soliton Solution:so(5)2}
\begin{align}
q_{12}^{+}(x,t)=  q_{45}^{+}(x,t)= & - \sum^{2}_{k=1}(\lambda^{+}_{k}-\lambda^{-}_{k}) R_k^{-1}(x,t)(n_{0,k}^{1}m_{0,k}^{2}  \re^{-\ri M_{k}^{+}(x,t)} +n_{0,k}^{4}m_{0,k}^{5} \re^{-\ri M_{k}^{-}(x,t)}), \nonumber
\\
q_{13}^{+}(x,t)= -q_{35}^{+}(x,t) = & -\sum^{2}_{k=1}(\lambda^{+}_{k}-\lambda^{-}_{k}) R_k^{-1}(x,t)(n_{0,k}^{1}m_{0,k}^{3}  \re^{-\ri M_{k}^{+}(x,t)} -n_{0,k}^{3}m_{0,k}^{5} \re^{-\ri M_{k}^{-}(x,t)}), \nonumber
\\
q_{14}^{+}(x,t)= q_{25}^{+}(x,t) =& -\sum^{2}_{k=1}(\lambda^{+}_{k}-\lambda^{-}_{k}) R_k^{-1}(x,t) (n_{0,k}^{1}m_{0,k}^{4}  \re^{-\ri M_{k}^{+}(x,t)} +n_{0,k}^{2}m_{0,k}^{5} \re^{-\ri M_{k}^{-}(x,t)}), \nonumber
\\
q_{21}^{-}(x,t)= q_{54}^{-}(x,t) = & -\sum^{2}_{k=1}(\lambda^{+}_{k}-\lambda^{-}_{k}) R_k^{-1}(x,t)(-n_{0,k}^{2}m_{0,k}^{1}  \re^{\ri M_{k}^{-}(x,t)}-n_{0,k}^{5}m_{0,k}^{4} \re^{\ri M_{k}^{+}(x,t)}), \nonumber
\\
q_{31}^{-}(x,t)= -q_{53}^{-}(x,t) = & -\sum^{2}_{k=1}(\lambda^{+}_{k}-\lambda^{-}_{k}) R_k^{-1}(x,t)(-n_{0,k}^{3}m_{0,k}^{1}  \re^{\ri M_{k}^{-}(x,t)} +n_{0,k}^{5}m_{0,k}^{3} \re^{\ri M_{k}^{+}(x,t)}), \nonumber
\\
q_{41}^{-}(x,t)= q_{52}^{-}(x,t) =& -\sum^{2}_{k=1}(\lambda^{+}_{k}-\lambda^{-}_{k}) R_k^{-1}(x,t) (-n_{0,k}^{4}m_{0,k}^{1}  \re^{\ri M_{k}^{-}(x,t)}-n_{0,k}^{5}m_{0,k}^{2} \re^{\ri M_{k}^{+}(x,t)}).\label{Two Soliton Solution:so(5)2a}
\end{align}
Here
\begin{align}
M_{k}^{\pm}(x,t)= &\lambda_{k}^{\pm}x + 2  f_{0, k}^{\pm} t,\qquad R_{k}(x,t)=2 R_{0,k} \rm {cosh} (2\nu_{k}x+2\tilde{\mu_{k}}t+ \xi_{0,k})+S,\label{Two Soliton Solution:so(5)2c}
\\
\nu_{k}= &\rm i (\lambda^{-}_{k}-\lambda^{+}_{k})/2, \quad  \tilde{\mu_{k}}= \rm i (\rm f(\lambda^{-}_{k})-\rm f(\lambda^{+}_{k})), \quad S= m_{0,k}^{2}n_{0,k}^{2}+ m_{0,k}^{3}n_{0,k}^{3}+ m_{0,k}^{4}n_{0,k}^{4},\label{Two Soliton Solution:so(5)2e}
\\
R_{0,k}= & \sqrt{  m_{0,k}^1 n_{0,k}^1  +  m_{0,k}^5 n_{0,k}^5 }, \qquad  \xi_{0,k}=\frac{1}{2}\rm {ln} \frac{ m_{0,k}^1 n_{0,k}^1 }{  m_{0,k}^5 n_{0,k}^5 }. \label{Two Soliton Solution:so(5)2f}
\end{align}
\end{subequations}
\end{example}

\begin{example} Again in the case when ${\frak g}\simeq so(5,{\Bbb C})$, if we impose a nonlocal involution of the form  \eqref{The Local Involutions2m1} then the dressing factor in \eqref{Two Soliton Solution:so(5)} will satisfy the following reduction condition:
\begin{align}\label{involution of two Soliton Solution:g=so(5)}
B \ \prod _{k=1}^{2} u_{k}(-x,t,-\lambda^{\ast})^{\dagger}\ B^{-1}=& \ \prod _{k=1}^{2} u^{-1}_{k}(x,t,\lambda),
\end{align}
where $B$ is constant block-diagonal matrix. This involution is satisfied if:
\begin{align}\label{involution of two Soliton Solution:g=so(5)1}
P_{k}(x,t)= & BP_{k}^{\dagger}(-x,t)B^{-1},\qquad (-\lambda^{\pm})^{\ast}=\lambda^{\mp}.
\end{align}
That means the projector $P_{k}(x,t)$ and $c_{k}(\lambda)$ become:
\begin{align}
P_{k}(x,t)= & \frac{|n_k(x,t)\rangle \langle n_k^{\ast}(-x,t)|B}{\langle n_k^{\ast}(-x,t)|B | n_k(x,t)\rangle}, \quad \langle m_k(x,t)|=(B | n_k(-x,t)\rangle)^{\dagger},\quad c_{k}(\lambda)= & \frac{\lambda-\lambda_{k}^{+}}{\lambda+(\lambda^{+}_{k})^{\ast}}.\label{involution of two Soliton Solution:g=so(5)2a}
\end{align}
So, from \eqref{One Soliton Solution:so(5)4}, it follows that
\begin{align}\label{involution of two Soliton Solution:g=so(5)3}
Q(x,t)=& Q_{0}(x,t)-(\lambda^{+}_{k}+(\lambda^{+}_{k})^{\ast}) [\rJ,BP_{k}^{\dagger}(-x,t)B^{-1}].
\end{align}
Thus,  the components of two soliton solution can be written as:
\begin{subequations}\label{involution of two Soliton Solution:g=so(5)4}
\begin{align}
q_{12}^{+}(x,t)= & q_{45}^{+}(x,t)= -\sum^{2}_{k=1}(\lambda^{+}_{k}+(\lambda^{+}_{k})^{\ast})R_k^{-1}(x,t)(n_{0,k}^{1} (n_{0,k}^{2})^{\ast} \re^{-\ri M_{k}^{+}(x,t)} -n_{0,k}^{4}(n_{0,k}^{5})^{\ast} \re^{-\ri (M_{k}^{+}(-x,t))^{\ast}}), \nonumber
\\
q_{13}^{+}(x,t)= &- q_{35}^{+}(x,t)= -\sum^{2}_{k=1}(\lambda^{+}_{k}+(\lambda^{+}_{k})^{\ast})R_k^{-1}(x,t)(n_{0,k}^{1} (n_{0,k}^{3})^{\ast}  \re^{-\ri M_{k}^{+}(x,t)} +n_{0,k}^{3}(n_{0,k}^{5})^{\ast} \re^{-\ri (M_{k}^{+}(-x,t))^{\ast}}), \nonumber
\\
q_{14}^{+}(x,t)= & q_{25}^{+}(x,t)= -\sum^{2}_{k=1}(\lambda^{+}_{k}+(\lambda^{+}_{k})^{\ast})R_k^{-1}(x,t)(n_{0,k}^{1} (n_{0,k}^{4})^{\ast}  \re^{-\ri M_{k}^{+}(x,t)} -n_{0,k}^{2} (n_{0,k}^{5})^{\ast} \re^{-\ri (M_{k}^{+}(-x,t))^{\ast}}), \nonumber
\\
q_{21}^{-}(x,t)= & q_{54}^{-}(x,t)= -\sum^{2}_{k=1}(\lambda^{+}_{k}+(\lambda^{+}_{k})^{\ast})R_k^{-1}(x,t)(n_{0,k}^{2} (n_{0,k}^{1})^{\ast}  \re^{\ri (M_{k}^{-}(-x,t))^{\ast}}-n_{0,k}^{5}(n_{0,k}^{4})^{\ast} \re^{\ri M_{k}^{+}(x,t)}), \nonumber
\\
q_{31}^{-}(x,t)= & - q_{53}^{-}(x,t)= -\sum^{2}_{k=1}(\lambda^{+}_{k}+(\lambda^{+}_{k})^{\ast})R_k^{-1}(x,t)(n_{0,k}^{3} (n_{0,k}^{1})^{\ast}  \re^{\ri (M_{k}^{-}(-x,t))^{\ast}} +n_{0,k}^{5}(n_{0,k}^{3})^{\ast} \re^{\ri M_{k}^{+}(x,t)}), \nonumber
\\
q_{41}^{-}(x,t)= & q_{52}^{-}(x,t)= -\sum^{2}_{k=1}(\lambda^{+}_{k}+(\lambda^{+}_{k})^{\ast})R_k^{-1}(x,t)(n_{0,k}^{4} (n_{0,k}^{1})^{\ast} \re^{\ri (M_{k}^{+}(-x,t))^{\ast}}-n_{0,k}^{5}(n_{0,k}^{2})^{\ast} \re^{\ri M_{k}^{+}(x,t)}).\label{involution of two Soliton Solution:g=so(5)4a}
\end{align}
Here,
\begin{align}
(M_{k}^{+}(-x,t))^{\ast}= &(-\lambda_{k}^{+})^{\ast}(-x) + 2 \ (\rm f(-\lambda^{+}_{1})^{*}) t, \quad R_{k}(x,t)=2 R_{0,k} \rm {cosh} (2\nu_{k}(x)+2\tilde{\mu_{k}}t+ \xi_{0,k})+S, \label{involution of two Soliton Solution:g=so(5)4b}
\\
\nu_{k}= & \rm i ((-\lambda^{+}_{k})^{*}-\lambda^{+}_{k})/2, \qquad  \tilde{\mu_{k}}= \rm i (\rm f(-\lambda^{+}_{k})^{*}-\rm f(\lambda^{+}_{k})),\label{involution of two Soliton Solution:g=so(5)4c}
\\
S= &(n_{0,k}^{2})^{*} n_{0,k}^{2}+ (n_{0,k}^{3})^{*} n_{0,k}^{3}+ (n_{0,k}^{4})^{*} n_{0,k}^{4},\label{involution of two Soliton Solution:g=so(5)4d}
\\
R_{0,k}= & \sqrt{  (n_{0,k}^1)^{*} n_{0,k}^1  +  (n_{0,k}^5)^{*} n_{0,k}^5 }, \qquad  \xi_{0,k}=\frac{1}{2}\rm {ln} \frac{ (n_{0,k}^1)^{*} n_{0,k}^1 }{  (n_{0,k}^5)^{*} n_{0,k}^5 }. \label{involution of two Soliton Solution:g=so(5)4e}
\end{align}
\end{subequations}

\end{example}

\section{Conclusions}\label{sec:6}

We have studied here multi-component generalisations of NLS models on {\bf A.III}  and {\bf BD.I} symmetric spaces. Our study is based on two types of main examples: the Manakov vector NLS equation, related to symmetric spaces of {\bf A.III}  type and Kulish-Sklyanin models, related to symmetric spaces of {\bf BD.I} type. Firstly, we formulated the direct scattering problem for both models. This includes: the construction of the Jost solutions, the scattering matrix and the minimal set of scattering
data. Based on the Gauss decomposition of the scattering matrix, we have also constructed the fundamental analytic solutions (FAS).

It turns out that the spectral properties of the Lax operator depend crucially on the choice of representation of the underlying Lie algebra
or symmetric space while the minimal set of scattering data is provided by the same set of functions \cite{GG2010}.

Finally, we have presented a modification of the dressing method and obtained 1- and 2-solitons for both models with nonlocal reductions.  Depending on the positions of the discrete eigenvalues $\lambda_j^\pm$ in the spectral plane, there are two regimes for the 2-soliton solution: if two of the discrete eigenvalues are in the upper half of the complex plane, while the other two are in the lower half, then the nonlocal involution will preserve their number inside each of the contours, so it will result in a Riemann-Hilbert problem with balanced number of singularities and therefore the corresponding 2-soliton solutions will be regular for all $t$. Otherwise, the 2-soliton solution will develop again a singularity in finite time as in the case, studied in \cite{AblMus,AblMus1,AblMus2}.

The results of this paper can be extended in several directions:
\begin{itemize}

\item To construct gauge covariant formulation of the multi-component NLS hierarchies on symmetric
spaces, including the Wronskian relations, the ``squared solutions'' and their completeness relations, the descriptions of the class of NLEEs associated to a given scattering problem, the generating (recursion) operator and it spectral decomposition, the description of the infinite set of
integrals of motion, the hierarchy of Hamiltonian structures and the $r$-matrix formulation.

\item To study the gauge equivalent systems of multi-component ferromagnetics  on symmetric spaces \cite{66,GGMV2011a,GGMV2011b}.

\item To study different types of reductions of multi-component integrable systems  on symmetric spaces \cite{gc1,gc,Metin}.

\item To study the associated Darboux transformations and their generalizations for NLS equations over
Hermitian symmetric spaces and to obtain multi-soliton solutions via such generalizations. This includes also  rational solutions \cite{DokLeb}.

\item To extend our results for the case of non-vanishing boundary conditions (a non-trivial
background) \cite{AblMus3,Li1}. The considerations required in this
case are more complicated and will be discussed it elsewhere.

  \item To study other types of integrable hierarchies on symmetric spaces, e.g. quadratic bundles (related to the derivative NLS equation, the Kaup-Newell equation or Gerdjikov-Ivanov equation) \cite{GGI2016,GGI2017,GVYa*08,Rei,ReiSTSH,78} or rational bundles \cite{GGMV2011a,GGMV2011b}.

\end{itemize}

\section*{Acknowledgements} The authors have the pleasure to thank Prof. Mark Ablowitz and Prof. Vladimir Gerdjikov for numerous useful discussions.

\end{document}